\documentclass[12pt]{iopart}
\usepackage{epsfig}
\newcommand{\p}{\partial}

\begin{document}

\begin{center}
\title{Magnetic Kronig-Penney model for Dirac
electrons in single-layer graphene}

\author{M.~Ramezani Masir}

\address{Departement Fysica, Universiteit Antwerpen Groenenborgerlaan 171, B-2020 Antwerpen, Belgium,}
\ead{mrmphys@gmail.com}

\author{P.~Vasilopoulos}

\address{Department of Physics, Concordia University, Montreal, Quebec,
Canada H3G 1M8.} \ead{takis@alcor.concordia.ca}

\author{F.~M.~Peeters}

\address{Departement Fysica, Universiteit Antwerpen Groenenborgerlaan 171, B-2020 Antwerpen, Belgium,\\
Departamento de F\'{\i}sica, Universidade Federal do Cear\'{a},
Caixa Postal 6030, Campus do Pici, 60455-760 Fortaleza, Cear\'{a},
Brazil} \ead{francois.peeters@ua.ac.be}

\end{center}
\begin{abstract}
The properties of Dirac electrons in a magnetic superlattice (SL) on
graphene consisting of  very high and thin ($\delta$-function)
barriers are investigated. We obtain the energy spectrum
analytically and study the transmission through a finite number of
barriers. The results are contrasted with those for electrons
described by the Schr\"{o}dinger equation. In addition, a
collimation of an incident beam of electrons is obtained along the
direction perpendicular to that of the SL. We also highlight the
analogy with optical media in which the refractive index varies in
space.
\end{abstract}
\maketitle
\section{Introduction}

During the last five years single-layer graphene (a monolayer of
carbon atoms) has become a very active field of research in
nanophysics \cite{no05,zh05}. It is expected that this material will
serve as a base for new electronic and opto-electric devices. The
reason is that graphene's electronic properties are drastically
different from those, say, of conventional semiconductors. Charge
carriers in a wide single-layer graphene behave like "relativistic",
chiral, and massless particles  with a "light speed" equal to the
Fermi velocity and possess  a {\it gapless,  linear} spectrum close
to the $K$ and $K'$ points. One  major consequence is  the perfect
transmission through arbitrarily high and wide barriers, referred to
as Klein tunneling.

One of the most challenging tasks is to learn how to control the
electron behavior using electric fields in graphene. This task is
made complicated precisely by the  Klein tunneling according to
which Dirac electrons in graphene can tunnel through arbitrarily
wide and high electric barriers  \cite{ka06}.

Alternatively,  one can apply a magnetic field to control the
electron motion. It was shown in numerous papers that an
inhomogeneous magnetic field  can confine standard electrons
described by the Schr\"{o}dinger equation \cite{pe93,Rei1,Rei2}. The
question then arises whether it can confine Dirac electrons in
graphene. Up to now semi-infinite magnetic structures, that are
homogeneous in one direction, were considered and made the task
simpler by converting the problem into an one-dimensional (1D) one
\cite{egg,rm08,mass,Park,Rakyta,Ghosh,zhai,MT,Xu}. In particular, a
magnetic confinement of Dirac electrons in graphene was reported in
structures involving one  \cite{egg} or several magnetic barriers
\cite{rm08,mass} as well as in superlattices, without magnetic field
for some very special values of the parameters involved
\cite{Cheol}. In such structures standard electrons can remain close
to the interface and move along  so-called snake orbits \cite{Rei1}
or in pure quantum mechanical unidirectional states \cite{pe93}.

\begin{figure}[ht]
\begin{center}
\includegraphics[width=10cm]{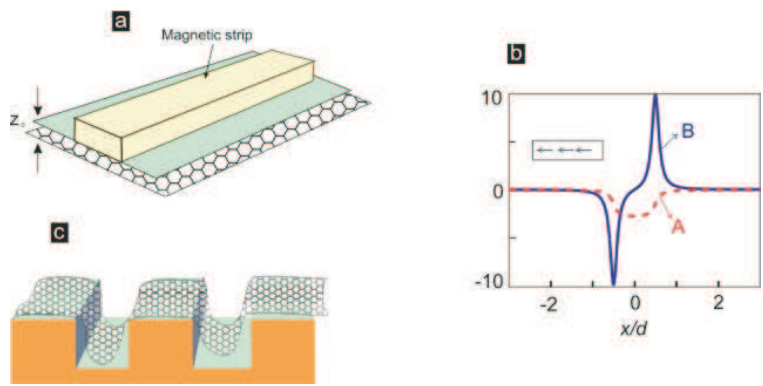}
\caption{(a) Layout of the system: a ferromagnetic stripe on top of
a bilayer graphene sheet separated by a thin oxide layer. (b)
Magnetic field and corresponding vector potential at a distance
$z_{0} = 0.1$ under the stripe for  parallel to it. (c) The graphene
layer on top of a periodic structured surface.}\label{fig0}
\end{center}
\end{figure}

Given the importance of graphene, it would be appropriate to study
this magnetic confinement  more systematically. We make  such a
study here by considering a {\it magnetic} Kronig-Penney (KP) model
in  graphene, i.e., a  series of magnetic $\delta$-function barriers
that alternate in sign. This  model can be realized experimentally
in two different ways:\\
1)  One can deposit ferromagnetic strips on top of a graphene layer
but in a way that there is no electrical contact between graphene
and these strips. When one magnetizes the strips along the $x$
direction, cf. Fig. \ref{fig0}(a), by, e.g., applying an in-plane
magnetic field, the charge carriers in the graphene layer feel an
inhomogeneous magnetic field profile. This profile can be well
approximated \cite{mat} by $2B_0z_0h/d(x^2+z_0^2)$  on one edge of
the  strip and by $-2B_0z_0h/(x^2+z_0^2)$ on the other, where $z_0$
is the distance between the 2DEG  and the strip, and $d$ and $ h$
the width and height of the strip (see Fig. \ref{fig0}(b)). The
resulting magnetic field profile will be modeled by two magnetic
$\delta$ functions of height $2\pi B_0 h$. Such ferromagnetic strips
were deposited on top of a  two-dimensional electron gas (2DEG) in a
semiconductor heterostructure in Ref. \cite{nog}.\\
2) It was recently shown that local strain in graphene induces an
effective inhomogeneous magnetic field \cite{Castro} (Fig.
\ref{fig0}(c)). When one puts the graphene layer on a periodically
structured substrate the graphene at the edges of the substrate
becomes strained and the situation can be described by a magnetic
$\delta$-function profile such as that shown in Fig. \ref{fig1}.

In a quantum mechanical treatment of the above two systems the
vector potential ${\bf A}(x)$ is the essential quantity and, within
the Landau gauge, ${\bf A}(x)$ is nothing else than a periodic array
of step functions. The Hamiltonian describing this system is
periodic and consequently we expect the energy spectrum of the
charge carriers in graphene to exhibit a band structure. The
advantage of this  {\it magnetic  Kronig-Penney} (KP) model is
mainly its analytical simplicity that provides some insight and
allows a contrast with the same model for standard electrons
\cite{ibra}. To do that we adapt a method developed in optics, for a
media with periodic in space refractive index. This optical method
is clear and very well suited to the problem. Incidentally, there
are many analogues of optical behavior in electronics, such as
focusing \cite{Van,Spec,Siv}, collimation or quasi-1D motion of
electrons and photons \cite{Cheol,Mole,pe93, rm08}, and interference
\cite{Yac} in a 2DEG.

The manuscript is organized as follows. In Sec. 2 we present the
method and evaluate the spectrum and electron transmission through
two antiparallel, $\delta$-function magnetic barriers.  In Sec. 3 we
consider superlattices of such barriers and present numerical
results for the energy spectrum. In Sec. 4 we consider a series of
$\delta$-function vector potentials and our concluding remarks are
given in Sec. 5.
\section{Characteristic matrix for Dirac electrons}
An electron in a single-layer graphene, in the presence of a
perpendicular magnetic field $B(x)$, which depends on  $x$, is
adequately described by the Hamiltonian
\begin{equation}\label{E1}
  H_0 = \displaystyle{ v_{F}\sigma\cdot( \textbf{p} +
  e\textbf{A}(x))},
\end{equation}
where ${\textbf{p}}$ is the momentum operator,  $v_{F}$ the Fermi
velocity, and ${\textbf{A}(x)}$ the vector potential. To simplify
the notation we introduce the dimensionless units: $ \ell_{B}
=[\hbar/ eB_{0}]^{1/2}, \, B(x) \rightarrow B_{0}B(x), \,\,{\bf
A}(x)\rightarrow B_{0} \ell_{B} {\bf A}(x), \,
    t\rightarrow t\ell_{B}/v_{F}, \,
    \vec{r} \rightarrow \ell_{B}\vec{r}, \,
    \vec{v} \rightarrow v_{F}\vec{v}, \,
    E \rightarrow E_{0} E,\,\, u(x) \rightarrow E_{0} u(x) ,\,\, E_{0}=\hbar v_{F}/\ell_{B}.$
Here $ \ell_{B}$ is the magnetic length and $t$ the tunneling
strength. In these units  Eq. (\ref{E1}) takes the form
\begin{equation}\label{E2}
    H=\left(
        \begin{array}{cc}
          0 & \p_{x} - i \p_{y} + A(x) \\
          \p_{x} + i \p_{y} - A(x) & 0 \\
        \end{array}
      \right).
\end{equation}
Then the equation $H\Psi(x,y)=E\Psi(x,y)$ admits  solutions of the
form
\begin{equation}\label{E3}
  \Psi(x,y) = \left(
                \begin{array}{c}
                  \psi_{I}(x,y) \\
                  \psi_{II}(x,y) \\
                \end{array}
              \right),
\end{equation}
with $\psi_{I}(x,y), \psi_{II}(x,y)$ obeying the coupled equations
\begin{eqnarray}
  \displaystyle{i\left[\frac{\p}{\p x} -i \frac{\p}{\p y} + A(x)\right]\psi_{II} + E \psi_{I}} &=& 0, \label{E4}\\
  \displaystyle{i\left[\frac{\p}{\p x} +i \frac{\p}{\p y}  -A(x)\right]\psi_{I} + E \psi_{II}} &=&
  0.\label{E5}
\end{eqnarray}

Due to the translational invariance along the $y$ direction we
assume solutions of the form $\Psi(x,y) = \exp{i k_{y}
y}(U(x),V(x))^{T}$, with the superscript $T$ denoting the transpose
of the row vector. For $B(x) \sim \delta(x)$ the corresponding
vector potential is a step function $A(x)\sim \Theta(x)$.
For $A(x)=P$ constant, Eqs. (\ref{E4}) and (\ref{E5}) take the form
\begin{eqnarray}
  \displaystyle{\left[\frac{d}{d x}  + (k_{y} + P) \right]V} &=& i E U ,\label{E6} \\
  \displaystyle{\left[\frac{d}{d x} - (k_{y} + P)\right]U }&=& i E V,\label{E7}
\end{eqnarray}
Equations (\ref{E3})-(\ref{E7})  correspond to those for an
electromagnetic wave propagating through a medium in which the
refractive index varies periodically. The two components of  $
\Psi(x,y)$  correspond to those of the electric (or magnetic) field
of the wave \cite{born,yar}.
Equations (\ref{E6}) and Eq. (\ref{E7}) can be readily decoupled by
substitution. The result is
\begin{equation}
  \displaystyle{\frac{d^2 Z}{d x^2}   + \left[E^{2} - (k_{y} + P)^{2}\right] Z} = 0,\label{E8} \\
\end{equation}
where $Z=U,V$. If $E^2\rightarrow E' $ and $(k_{y} + P)^2\rightarrow
V_{eff}$, Eq. (\ref{E8}) reduces to a Schr\"{o}dinger equation for a
standard electron where $V_{eff}(k_{y},x) = (k_{y}+P)^{2}$ can be
considered as an effective potential. Taking $\theta_{0} $ as the
angle of incidence, we have $k_{x} = E \cos{\theta_{0}}=[E^{2} -
k_{y}^2]^{1/2}$ and $k_{y} = E \sin{\theta_{0}}$ are the wave vector
components outside the medium and $k'_{x} = E \cos{\theta}=[E^{2} -
(k_{y} + P)^2]^{1/2}$ is the electron wave vector inside the medium
and $\theta=\tan^{-1}(k_{y}/k'_{x})$ is the refraction angle. This
renders Eq. (\ref{E8}) simpler with acceptable solutions for $U$ and
$V$
\begin{equation}\label{E9}
    \displaystyle{U(x) = A \cos\left(E x \cos{\theta} \right) + B \sin\left(E x \cos{\theta} \right)},
\end{equation}
\begin{equation}\label{E10}
    \displaystyle{V(x) = -i\left\{B\cos\left(\theta + E x \cos{\theta}
    \right) - A \sin\left(\theta  + E x \cos{\theta}\right)\right\}}.
\end{equation}
For future purposes, we write $U$ and $V$ as a linear combination of
$U_{1}, U_{2}$ and $V_{1},V_{2}$:
\begin{equation}\label{E11}
\begin{array}{cc}
    \displaystyle{\frac{d V_{1}}{d x}+(k_{y} + P)V_{1} = i E U_{1},} & \quad \displaystyle{ \frac{d V_{2}}{d x}+(k_{y} + P)V_{2} = i E U_{2}.}\\
    \\
  \displaystyle{\frac{d U_{1}}{d x}-(k_{y} + P)U_{1} = i E V_{1},} & \quad \displaystyle{ \frac{d U_{2}}{d x}-(k_{y} + P)U_{2} = i
  E V_{2}.}
\end{array}
\end{equation}
We now multiply the equations of the first row by $U_2$ and $U_1$,
respectively, and those of the second by $V_2$ and $V_1$.  The
resulting equations lead to
\begin{equation}\label{E12}
    \displaystyle{ \frac{d  \textbf{D}}{d x}
    } = U'_{1}V_{2} +U_{1}V'_{2}-
    V'_{1}U_{2}-V_{1}U'_{2}=0,
\end{equation}
where $ \textbf{D}=det D$ and

\begin{equation}\label{E13}
D = \left(
    \begin{array}{cc}
      U_{1}& V_{1}\\
      U_{2} & V_{2}
    \end{array}
    \right).
\end{equation}
Equation (\ref{E12})  shows that the determinant of the matrix
(\ref{E13}) associated with any two arbitrary solutions of Eq.
(\ref{E8}) is a constant, i.e, $D$ is an invariant of the system of
Eqs. (\ref{E11}). This also follows from the well-known property of
the Wronskian of second-order differential equations. For our
purposes the most convenient choice of  particular solutions is
\begin{equation}\label{E14}
    \begin{array}{cc}
      U_{1} = f(x), &\quad \quad  U_{2} = F(x),\\
      V_{1} = g(x), &\quad \quad  V_{2} = G(x),
    \end{array}
\end{equation}
such that
\begin{equation}\label{E15}
    \displaystyle{f(0)=g(0)=0,  
    \quad \quad F(0) = G(0) = 1}.
\end{equation}
Then the solution with $U(0)=U_{0} $, \, $V(0) = V_{0}$, can  be
expressed as
\begin{equation}  \label{E16}
       \displaystyle{U = F U_{0} + f V_{0}, \,\,\, V = G U_{0} + g V_{0}}
\end{equation}
or, in matrix notation, as
\begin{equation}\label{E17}
    \begin{array}{ccc}
\hspace*{-0.45cm}           \textbf{Q} = \left[
    \begin{array}{c}
      U(x) \\
      V(x)
    \end{array}
    \right], &     \textbf{Q}_{0} = \Big[
    \begin{array}{c}
      U_{0} \\
      V_{0}
    \end{array}
    \Big], &    \textbf{N} = \left[
    \begin{array}{cc}
      F(x)& f(x)\\
      G(x)& g(x)
    \end{array}
    \right].
    \end{array}
\end{equation}
%
\begin{figure}[ht]
\begin{center}
\includegraphics[width=10cm]{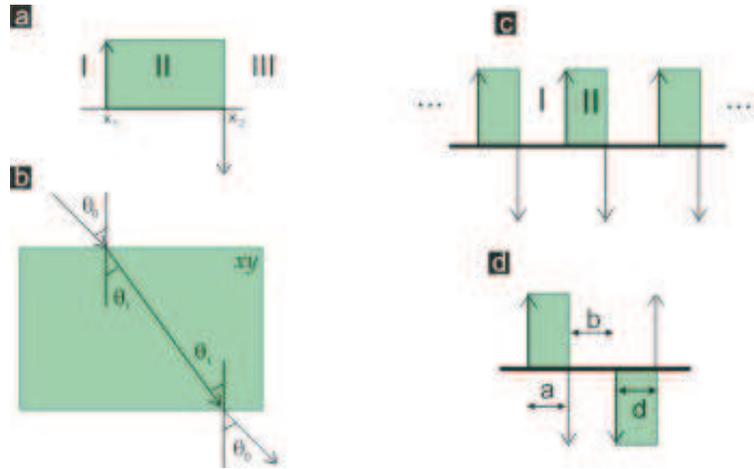}
\caption{(a) Two opposite magnetic $\delta$-function barriers,
indicated by arrows (top figure) the vector potential showed by
shaded (green) area. (b) The angles related to the propagation of an
electron through this system (is shown in the bottom figure). (c)
Schematics of a periodic vector potential (shaded areas) and
corresponding magnetic field indicated by the black arrows. (d)
Arrangement of magnetic $\delta$-functions that leads to a periodic
area of vector potential with alternating sign.}
\end{center}
\end{figure}\label{fig1}
Since $D $ is  constant, the determinant of the square matrix $N$ is
a constant; its value, found by setting $x = 0$,  is $ \det N = Fg -
fG = 1$.
It is usually more convenient to express $U_{0}$ and $V_{0}$ as a
function of $U(x)$ and $V(x)$. Solving for $U_{0}$ and $V_{0}$  we
obtain $ \displaystyle{\textbf{Q}_{0} = \textbf{M}\textbf{Q}}$,
where
\begin{equation}\label{E18}
    \textbf{M} = \left[
    \begin{array}{cc}
      g(x)& -f(x)\\
      -G(x)& F(x)
    \end{array}
    \right].
\end{equation}
This matrix M is unimodular, $|\textbf{M}| = 1$.
Now we can find the characteristic matrix from Eqs. (\ref{E9}) and
(\ref{E10}) as
\hspace*{-0.45cm} \begin{equation}\label{E19}
    \textbf{M}(x) = \frac{1}{\cos{\theta}}\left[
    \begin{array}{cc}
      \cos\left(\theta + E x \cos{\theta} \right)& \displaystyle{-i\sin\left(E x \cos{\theta} \right)} \\
      -i\sin\left( E x \cos{\theta} \right) & \cos\left(\theta - E x \cos{\theta}\right)
    \end{array}
    \right].
\end{equation}
\subsection{Bound states}
Regards to the average of vector potential we shall consider two
different systems: one with zero average and the other with non zero
average along the $x$-direction.
\begin{figure}[ht]
\begin{center}
\includegraphics[width=10cm]{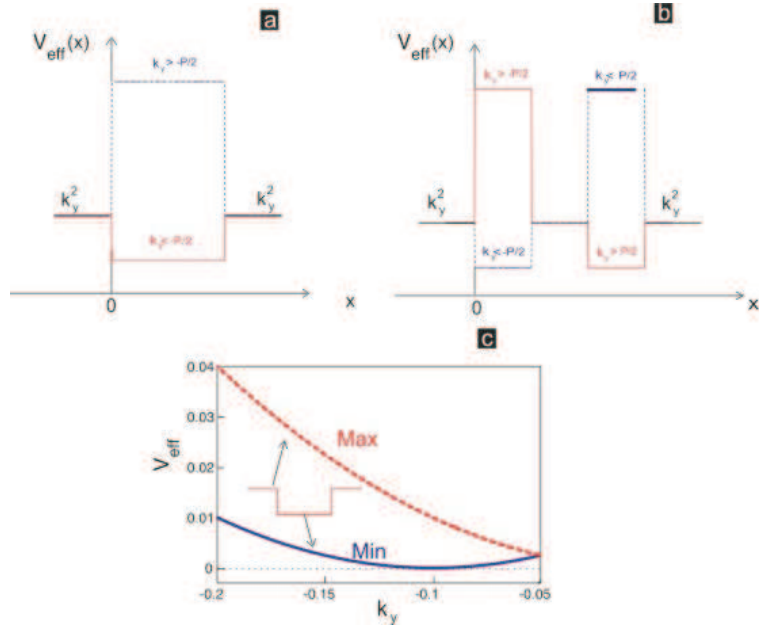}
\caption{The effective potential for $k_{y}<-P/2$ and $k_{y}>P/2$
for two different cases, (a) vector potential with non-zero average
corresponding to Fig. \ref{fig1}(a), and (b) vector potential with
zero average corresponding to Fig. \ref{fig1}(c). (c) Minimum (blue
full curve) and maximum (red dashed curve) of the effective
potential vs $k_{y}$ corresponding to the situation depicted in (a)
for $P = 0.1$. }\label{fig2a}
\end{center}
\end{figure}
First let us consider the magnetic field profile as shown in Fig.
{\ref{fig1}}(a) for which the corresponding vector potential is
\begin{equation}\label{A1}
    A(x) = P\Theta(x)\Theta(L-x),
\end{equation}
where $\Theta(x)=0(x<0)$, $1(x>0)$ is the theta function. This
vector potential has a non-zero average, and the corresponding
effective potential becomes (see Fig. \ref{fig2a}(a)) as
\begin{equation}\label{V1}
    V_{eff}(k_{y},x) \sim \left\{\begin{array}{l}
                    k_{y}^{2} \quad \quad \quad \quad\quad \quad x < 0\\
                    (k_{y} + P)^{2}  \quad \quad 0<x < L\\
                    k_{y}^{2} \quad \quad\quad \quad \quad \quad x > L
                  \end{array}
                  \right..
\end{equation}
Here $L$ is measured in the unit of magnetic length $l_{B}$. There
are two different cases which we have to consider.

\emph{Case} 1 for $k_{y}<-P/2$: as shown in Fig. \ref{fig2a}(a) by
the full red curve, we have a 1D symmetric quantum well which, as is
well-known, has at least one bound state (see also Fig.
\ref{fig2a}(c)). For $E^{2} < k_{y}^{2} $ the particle will be bound
while for $E^{2} > k_{y}^{2} $ we have scattered states, or
equivalently the electron tunnels through the magnetic barriers.

\emph{Case} 2 for $k_{y}> - P/2$: as shown in Fig. \ref{fig2a}(a) by
the dotted blue curve, the effective potential is like a barrier. We
have a pure tunneling problem. With reference to Fig. \ref{fig1},
$x_1=0$ and $x_2=L$, the solutions are as follows. For $x<0$ the
wave function is
\begin{equation}\label{E33}
    \psi(x) = C e^{\kappa x}\Big(
                                     \begin{array}{c}
                                       1 \\
                                       -i\e^{-\xi} \\
                                     \end{array}\Big),
\end{equation}
where $\kappa =E \cosh{\xi}$ and $k_{y} = E \sinh{\xi}$, while for
$x>L$ it is
\begin{equation}\label{E34}
    \psi(x) = De^{-\kappa x}\Big(
                                     \begin{array}{c}
                                       1 \\
                                       i \e^{\xi} \\
                                     \end{array}\Big).
\end{equation}
%
In the middle region, $0 < x < L $, the wave function is given by
\begin{equation}\label{E35}
    \psi(x) =  Fe^{ik'x}\Big(
                                     \begin{array}{c}
                                       1 \\
                                       e^{i \theta} \\
                                     \end{array}\Big)+
                                   Q e^{-ik'x}\Big(
                                     \begin{array}{c}
                                       1 \\
                                       -e^{-i \theta} \\
                                     \end{array}\Big).
\end{equation}
With $k' =[E^{2} -(k_{y} + P)^{2}]^{1/2} = E\cos{\theta}$. Matching
the wave functions at $x=0$ and $x = L$ leads to a system of four
equations relating the coefficients $C$, $ D$, $F$, and $ Q$.
Setting the determinant of these coefficients equal to zero, we
obtain the transcendental equation, the solution of it gives the
energy spectrum
\begin{equation}\label{E36}
    \cos{\theta}\cosh{\xi}\cos{k'L} + \sin{\theta}\sinh{\xi}\sin{k'L}=0.
\end{equation}
%
\begin{figure}[ht]
\begin{center}
\includegraphics[width=10cm]{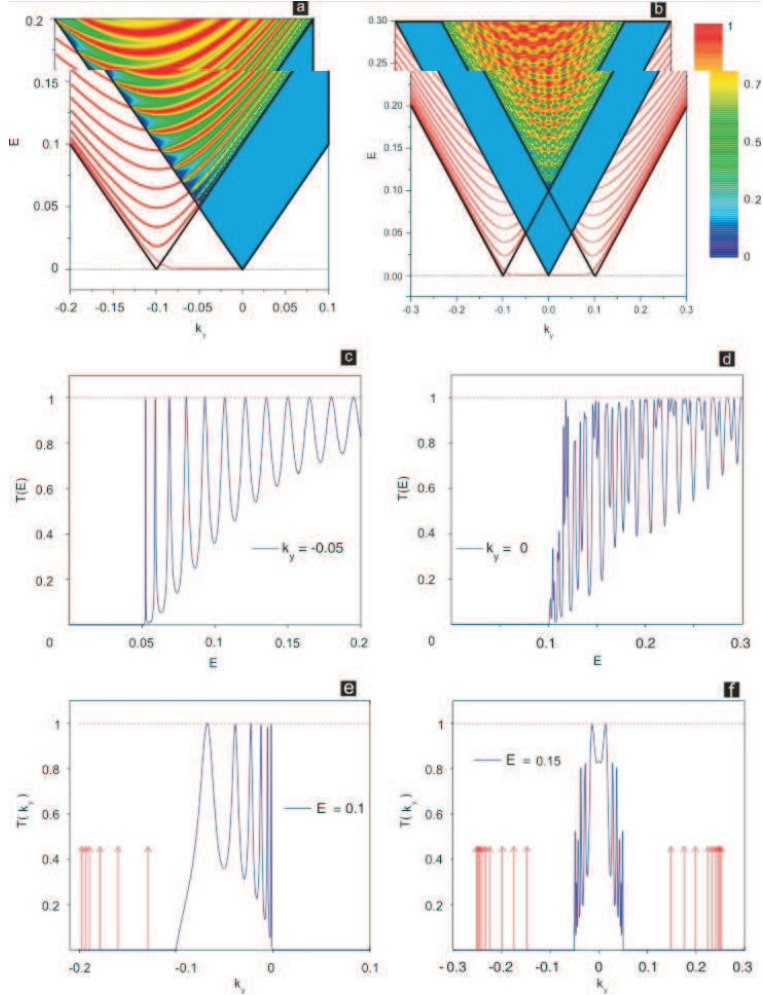}
\caption{(a) Energy spectrum (red full curves on white background)
and contour plot of the transmission  (color  background) through
two magnetic, but with opposite direction  $\delta$-function
barriers for $L=200$ and $P = 0.1$. (b) the same as (a) but now for
the configuration shown in Fig. \ref{fig1}(b) for $a = b = c= 200$
and $P = 0.1$. (c) Transmission vs energy through the magnetic
$\delta$-function barriers of in (a) $k_{y}=-0.05$. (d) same as (c)
but now for the configuration shown in Fig. \ref{fig1}(b) for
$k_{y}=0$. (e) Transmission vs $k_{y}$ through the system described
in (a) for fixed $E=0.1$. The red arrow lines indicate the position
of the bound states. (f) same as (e) but now for the configuration
shown in Fig. \ref{fig1}(b) for $E = 0.15$}\label{fig2}
\end{center}
\end{figure}
For the special value of $k_{y}=-P$, $\sin{\theta}=0$, and we can
rewrite Eq. (\ref{E36}) as $\cos{(EL)}=0$ or equally
$\displaystyle{E_{n} = \left(n + \frac{1}{2}\right)\frac{\pi}{L}}$.
The resulting bound states, as a function of $k_y$, are shown by the
red full curves in Fig. \ref{fig2}(a).
The area of existence of bound states is delimited by the lines $E =
-k_{y}$ and $E = -(k_{y} + P)$. The number of bound states increases
with $|k_{y}|$ which is also clear from the behavior of the min. and
max. of the effective potential (see Fig. \ref{fig2a}(c)). No bound
states are found for $k_{y}>-P/2$ as is also apparent from Fig.
\ref{fig2a}(c). For $k_{y}\rightarrow-P/2$ the potential is shallow
and only one bound state exists.
The average velocity $v_{n}(k_{y})$ along the $y$ direction is given
by
\begin{equation}\label{V2}
         v_{n}(k_{y}) = \p E/\p k_{y} =
       \int^{+\infty}_{-\infty}dx\,j_{y}(x),
\end{equation}
where $j_{y} = -i(U^{*}V-V^{*}U)$. From Fig. \ref{fig2}(a) it is
clear that these bound states move along the $y$-direction, i.e.
along the magnetic barriers. Their velocity $v_{y}>-v_{F}$ is
negative for $k_{y}< -P$ but as the electron is approaching $k_{y}
\rightarrow-P$ we have $v_{y}\rightarrow 0$. For $k_{y}>-P$ the
velocity $v_{y}>v_{F}$ is positive. This can be understood from the
maximum and minimum of the effective potential which is shown in
Fig. \ref{fig2a}(c). The energy bound states can only exist between
these two lines. Notice that the slope of min $V_{eff}$ is negative
for $k_{y}<-P$ while it turns positive for $k_{y}>-P$ which explains
the $k_{y}$ dependence of the velocity. From Fig. \ref{fig2}(a) it
is clear there are two different classes of bound states. The bound
state which follows very closely the $k_{y} = -P$ curve and extends
to the region $-P<k_{y}<0$ with energy close to zero has a
wavefunction that is concentrated around the position of the two
magnetic delta function and decays exponentially in the region
$0<x<L$. The wave function of the other bound states are
concentrated in a region between the two magnetic delta-functions
(i.e. like in a standing wave fashion) and decays exponentially
outside this region.

Next, we consider a structure with zero-average vector potential as
shown in Fig. \ref{fig1}(b), with corresponding effective potential
shown in Fig. \ref{fig2a}(b). The effective potential for
$k_{y}<-P/2$ and $k_{y}>P/2$, consist of a potential well and a
potential barrier and therefore has at least one bound state. Thus
we expect bound states for all $k_{y}$ with energy between $E =
-(+)(k_{y} + P)$ and $E = -(+)k_{y}$ when $k_{y}<-P/2$
($k_{y}>P/2$). The dispersion relation for those bound states are
the solution of
\begin{equation}\label{E36d}
    M_{21} - M_{12} - iM_{22}e^{-\xi} - iM_{11}\e^{\xi} =0,
\end{equation}
where $\textbf{M}$ is the transfer matrix for the unit shown in Fig.
\ref{fig1}(d). These bound states are shown by the red full curves
in Fig. \ref{fig2}(b). Because of the spatial inversion symmetry of
the vector potential the spectrum has the symmetry
$E(-k_{y})=E(k_{y})$. Notice that for $-P<k_{y}<P$ the lowest bound
state has energy $E \approx 0$. For $-P/2<k_{y}<P/2$ we have two
potential barriers and therefore no bound states.
\subsection{Reflection and transmission coefficients}
Consider a plane wave incident upon a system of two
$\delta$-function magnetic barriers, identical in height but
opposite in direction, placed at $x = 0$ and $x = L$, as shown
schematically in Fig. \ref{fig1}(a). In this case the vector
potential is constant for $ 0\leq x \leq L$, zero outside  this
region, and homogeneous in the $y$ direction. Below we derive
expressions for the amplitudes and intensities of the reflected and
transmitted waves.

Let $A$, $R$, and $T$ denote the amplitudes of the incident,
reflected, and transmitted waves, respectively. Further, let
$\theta_{0}$ be the angle of incidence and exit as shown in Fig.
\ref{fig1}(b).
 The boundary conditions give
\begin{equation}\label{E20}
    \begin{array}{cc}
      U_{0} = A + R, \quad \quad& U(L) = T\e^{i k L}, \\
      V_{0} = A \e^{i\theta_{0}} - R \e^{-i\theta_{0}},\quad & V(L) = \e^{i\theta_{0}}
      \e^{i k h} T.
    \end{array}
\end{equation}
The four quantities $U_{0},V_{0}, U$, and $V$ given by Eqs.
(\ref{E20}) are connected by the basic relation $\textbf{Q}_{0} =
\textbf{M}\textbf{Q}$;  hence, with $ J=m'_{11} +
m'_{12}\e^{i\theta_{0}}$ and $ K=m'_{21} + m'_{22}\e^{i\theta_{0}}$,
we have
\begin{equation}\label{E21}
    \begin{array}{cc}
      A + R = J T \e^{i k L}, \\
      A \e^{i\theta_{0}} - R \e^{-i\theta_{0}}= K T\e^{i k L},
    \end{array}
\end{equation}
where $m'_{ij}$ are the elements of the characteristic matrix of the
medium, evaluated at $x = L$.  From Eq. (\ref{E21}) we obtain the
reflection and transmission amplitudes 
\begin{equation}\label{E22}
    r =\displaystyle{\frac{R}{A} = \frac{J \e^{i\theta_{0}}-K}{J\e^{-i\theta_{0}}+ K}},
 \quad\quad   t =\displaystyle{\frac{T}{A} = \frac{2\e^{-i k h}\cos{\theta_{0}}}{J  \e^{-i\theta_{0}}+K}}.
\end{equation}
In terms of $r$ and $t$ the \emph{reflectivity }and
\emph{transmissivity} are
\begin{equation}\label{E23}
    \begin{array}{cc}
    \mathcal{R} = |r|^{2}, \quad \quad & \mathcal{T} =|t|^{2}.
    \end{array}
\end{equation}
The characteristic matrix for a homogeneous vector potential is
given by Eq. (\ref{E19}). Labeling with subscripts 1, 2, and 3
quantities which refer to the regions, respectively, I, II, and III
of Fig. \ref{fig1}(a), and by $L=x_{2} -x_{1}$ distance between the
magnetic $\delta$-functions, we have ($ \beta=\displaystyle{E
L\cos{\theta_{i}}}$)
\begin{eqnarray}
\label{E24} \nonumber
&\hspace*{-0.35cm} m'_{11} = \displaystyle{\cos\left(\theta_{i} + \beta \right)/\cos{\theta_{i}}},\,\,\, m'_{22} =\displaystyle{\cos \left(\theta_{i} - \beta \right)/\cos{\theta_{i}}}, \\
& \hspace*{-0.95cm}  m'_{12} = \displaystyle{-i\sin
\beta/\cos{\theta_{i}}}, \,\,\,\,m'_{21} =\displaystyle{-i \sin
\beta/\cos{\theta_{i}}}.
\end{eqnarray}
The reflection and transmission amplitudes $r$ and $t$ are obtained
by substituting these expressions in those for $J$ and $K$ that
appear in Eq. (\ref{E22}). The resulting formula can be expressed in
terms of the
amplitudes $r_{12},t_{12}$ and $r_{23},t_{23}$ associated with the
reflection at and transmission through the first and second
"interface", respectively. We have
\begin{equation}\label{E25}
    r_{12} = \displaystyle{\frac{\e^{i\theta_{0}}-\e^{i\theta_{i}}}{\e^{-i\theta_{0}}+ \e^{i\theta_{i}}}},
%
 \quad  \quad  t_{12} = \displaystyle{\frac{2\cos{\theta_{0}}}{{\e^{-i\theta_{0}}+ \e^{i\theta_{i}}}}},
\end{equation}
and similar expressions for $r_{23}$ and $t_{23}$. In terms of
these expressions 
$r$ and $t$ become
\begin{equation}\label{E26}
    r = \displaystyle{\frac{r_{12}+ r_{23}\e^{2 i \beta}}{1 + r_{12}r_{23}\e^{2 i \beta}}},
  \quad  \quad   t = \displaystyle{\frac{t_{12}t_{23}\e^{i \beta}}{1 + r_{12}r_{23}\e^{2 i \beta}}}.
\end{equation}
The amplitude $t$ of the transmission  through the system   is given
by \cite{rm08,mass,Ghosh,Castro},
\begin{equation}\label{E27}
t = \frac{2\e^{-i k L} \cos{\theta_{0}}\cos{\theta_{i}}}
{e^{-i\beta}[\cos(\theta_{0} +
\theta_{i})+1]+e^{i\beta}[\cos(\theta_{0} - \theta_{i})-1]},
\end{equation}
where $k_{y} = E \sin{\theta_{0}}$, and $k_{y} + P = E
\sin{\theta_{i}}$. This equation remains invariant under the changes
$E\rightarrow -E$ , $\theta_{0}\rightarrow -\theta_{0},
\theta_{i}\rightarrow -\theta_{i}$. A contour plot of the
transmission is shown in Fig. \ref{fig2}(a) and slices for constant
$k_{y}$ and $E$ are shown respectively in Fig. \ref{fig2}(c) and
Fig. \ref{fig2}(d).
By imposing the condition that the wave number $k_{x}$  be real for
incident and transmitted waves, we find that the angles $\theta_{0}$
and $\theta_{i}$ are related by
\begin{equation}\label{E28}
    \displaystyle{\sin{\theta_{0}} + P/E = \sin{\theta_{i}}}.
\end{equation}
Equation ({{\ref{E28}}) expresses the angular confinement of the
transmission elaborated in Refs. \cite{rm08}, \cite{Ghosh},
\cite{Dell}, and \cite{mass}. Notice its formal similarity with
Snell's law. Using Eq. ({\ref{E28}}) we obtain the range of
incidence angles $\theta_{0}$ for which transmission through the
first magnetic barrier is possible
\begin{equation}\label{E30}
    \begin{array}{c}
      \displaystyle{-1 - P/E \leq \sin{\theta_{0}}\leq 1 - P/E}. \\
    \end{array}
\end{equation}
%
For the special value of the energy $\displaystyle{E = P/2}$ and
$\theta_{0}$ in the range $\displaystyle{-\pi/2\leq \theta_{0}\leq
\pi/2}$, we have $\theta_{i} = \displaystyle{\pi/2} $ while for
$\displaystyle{E = -P/2}$ the result is $\theta_{i} =
\displaystyle{-\pi/2} $. Alternatively, we can put $\theta_{i} = \pm
\displaystyle{\pi/2}$ in Eq. ({\ref{E28}}) and obtain, for $P>0$,
the result
\begin{equation}\label{E31}
    \displaystyle{\sin\theta^{\pm}_{0}= \pm1 -P/E},
\end{equation}
where the  $+ (-)$ sign corresponds to $E >0$ ($E <0$). A contour
plot of the transmission as function of $E$ and $k_{y}$, obtained
from Eq. (\ref{E27}), is shown in Fig. {\ref{fig2}}(a). In Fig.
{\ref{fig2}}(a) we distinguish three different regions. In the
region between $E = -(k_{y}+P/2)$ and $E =-k_{y}$, the wave vector
of the incident wave is imaginary and they are evanescent waves. In
this region $k'$ is real and it is possible to find localized
states. The $k$ and $k'$ for the second region between $E =
k_{y}+P/2$ and $E =-(k_{y}-P/2)$ are real and the electron can
tunnel through the magnetic $\delta$-barriers. In the blue shadow
region between  $E = k_{y}+P/2$ and $E = k_{y}-P/2$, $k$ is real but
$k'$ is imaginary and solutions inside the barrier are evanescent
and there is very little tunneling which becomes very quickly zero.
The transmission probability $|T| = t\cdot t^{\ast}$ is equal to $1$
for $\cos{(2\beta)} = 1$. In this case the energy becomes
\begin{equation}\label{E32}
    \displaystyle{E_n = \pm \Big[n^{2} \pi^{2}/L^{2} + (k_{y} +
    P)^{2}\Big]^{1/2}, \quad \quad \quad \quad \quad n = 1,2,...}.
\end{equation}
The condition $\cos{(2\beta)} = 1$, or equivalently $\beta=n\pi =
Ehcos\theta_2$ with $n$  an integer, should be combined with that
for the transmission to occur in the region delimited by the curves
$E = \pm(k_{y} + P)$ and $E = \pm k_{y}$. For example, in Fig.
\ref{fig2}(a) for $k_{y} = -0.05$ and $0 < E < 0.2$ we have $12$
maxima. It is readily seen that  with these parameters in Eq.
(\ref{E32}) we find $12$ different energies as  shown in Fig.
\ref{fig2}(c).\\
Fig. \ref{fig2}(b) shows a contour plot of the transmission for the
structure shown in Fig. \ref{fig1}(d), which is symmetric around
$k_{y}=0$. Notice that the number of resonances has increased
substantially as compared to previous case which is due to the fact
that we have twice as many magnetic barriers in our systems.
\section{A series of units with magnetic $\delta$-function barriers}
\subsection{$N$ units}
We consider a system of $N$ units, such as those shown in Fig.
\ref{fig1}(a) and Fig. \ref{fig1}(d) with periods $L = a + b$ and $L
= a + b + c + d$, respectively. The corresponding periodic vector
potential is $\textbf{A}(x)=\textbf{A}(x + nL)$  and the magnetic
field $\textbf{B}=\textbf{B}(x + nL) $, with $n=1,2,...,N$. The
characteristic matrix for one period $\textbf{M}(L)$ is denoted by
\begin{equation}\label{E37}
    \textbf{M}(L) = \left[
    \begin{array}{cc}
      m_{11}& m_{12} \\
      m_{21}& m_{22}
    \end{array}
    \right].
\end{equation}
On account of the periodicity we have
\begin{equation}\label{E38}
    \displaystyle{\textbf{M}(NL) = \underbrace{\textbf{M}(L)\cdot\textbf{M}(L)...\textbf{M}(L)}_{N factors} = (\textbf{M}(L))^{N}}.
\end{equation}
To evaluate the elements of $\textbf{M}(NL)$ we use a result from
the theory of matrices, according to which the \emph{N}th power of a
unimodular matrix $\textbf{M}(L)$ is ($u_N(\chi)\equiv u_N$)
\begin{equation}\label{E39}
\hspace*{-0.15cm}     \left[\textbf{M}(L)\right]^{N} = \left[
    \begin{array}{cc}
      m_{11}u_{N-1} - u_{N-2}& m_{12}u_{N-1} \\
      m_{21}u_{N-1}& m_{22}u_{N-1}-u_{N-2}
    \end{array}
    \right],
\end{equation}
with $ \displaystyle{\chi = \frac{1}{2}\Tr{\textbf{M}}}$ and $u_{N}$
the Chebyshev polynomials of the second kind:
\begin{equation}\label{E40}
    \displaystyle{u_{N}(\chi) = \sin[(N+1)\zeta]/\sin{\zeta}},
\end{equation}
where
\begin{equation}\label{E40a}
    \displaystyle{\zeta =  \cos^{-1}{\chi}},
\end{equation}
Here $\zeta$ is the \emph{Bloch phase} of the periodic system
\cite{Griff}, which is related to the eigenfunctions of
$\textbf{M}$. In the limit case of $N \rightarrow \infty$, we have
total reflection when $\zeta $ is outside the range $(-1,1)$.
\begin{figure}[ht]
\begin{center}
\includegraphics[width=10cm]{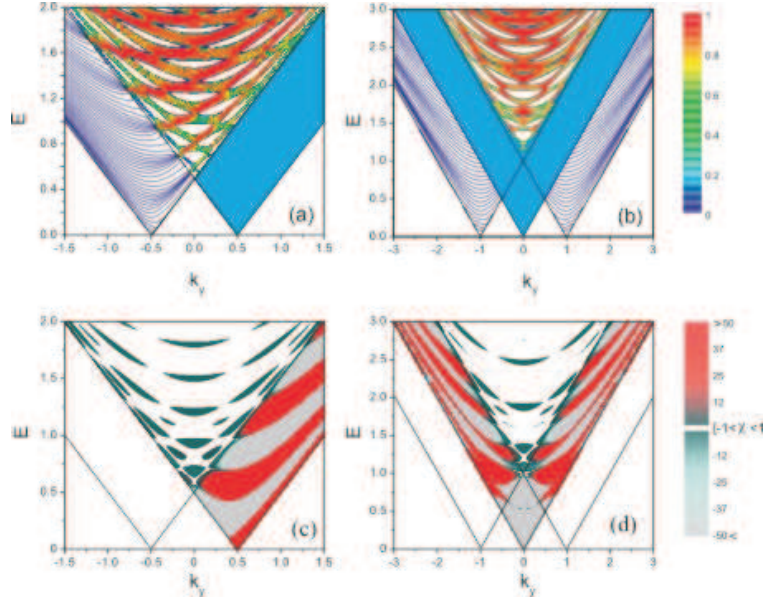}
\caption{Contour plot of the transmission (a) and Bloch phase (c)
through $N=10$ magnetic $\delta$-function barriers with $a = 10$, $b
= 10$, and $P = 1$. (b) and (d) The same as in (a) and (c) for $a =
5$, $b = 5$, $c = 5$, $d = 5$, and $P = 1$, single unit.}
\label{fig3}
\end{center}
\end{figure}
\begin{figure}[ht]
\begin{center}
\includegraphics[width=10cm]{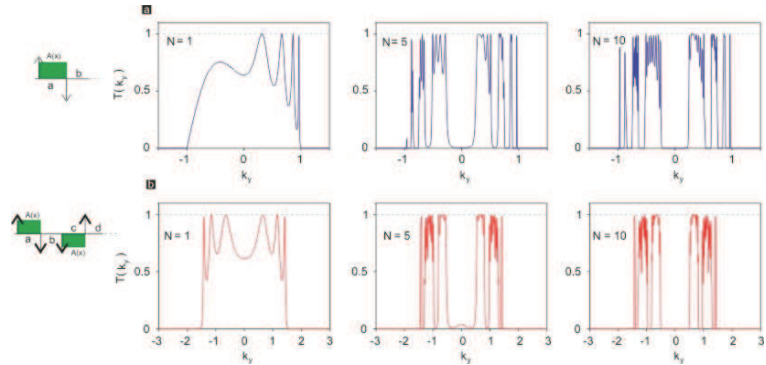}
\caption{(a) and (b) Transmission vs energy through  $N=1, 5, 10$
magnetic units of $\delta$-function barriers shown on the left. The
upper unit has $a = 10$, $b = 10$, $P = 1$, $E = 1.5$ and the bottom
one $a = b = c = d = 5$, and $P = 1$ and $E = 2.5$.} \label{fig4}
\end{center}
\end{figure}
%
\subsection{Superlattice}
Here we consider a finite number $N$ of lattice unit shown in Fig.
\ref{fig1}(c). We set
\begin{equation}\label{E41}
    \begin{array}{c}
      \beta_{2} = E b \cos{\theta_{2}}, \,   \beta_{1} = E a \cos{\theta_{1}},
 \,   p_{2} = \displaystyle{1/\cos{\theta_{2}}},\, \\   p_{1} = \displaystyle{1/\cos{\theta_{1}}}, \,\,
      h = a + b,\quad   \lambda^{\pm}_{n} = \theta_{n}\pm
      \beta_{n}.
    \end{array}
\end{equation}
The characteristic matrix $\textbf{M}_{2}(L)$ for one period is
readily obtained, in terms of these quantities, as in Sec. II, and
from that the characteristic matrix $\textbf{M}_{2N}(NL)$ of the
multilayer system according to Eq. (\ref{E38}). Its elements are
\begin{figure}[ht]
\begin{center}
\includegraphics[width=10cm]{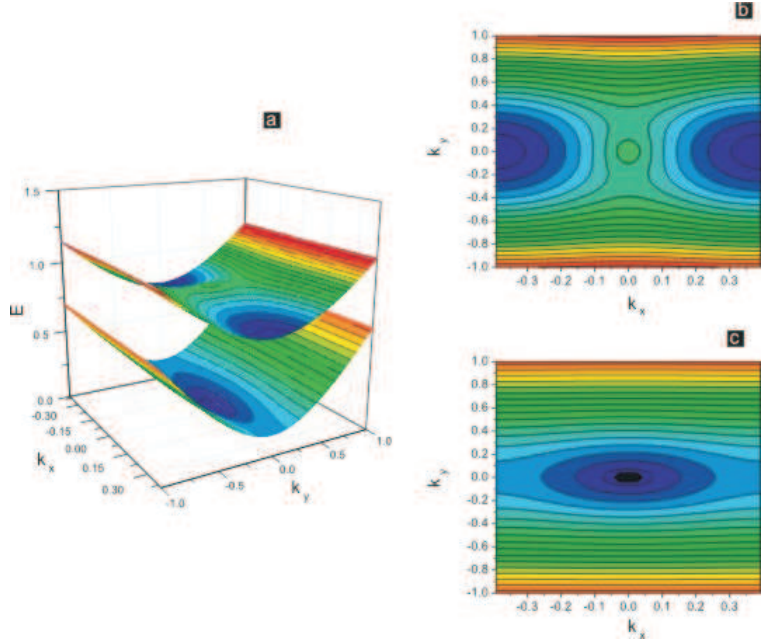}
\caption{(a) The first two energy bands and contour plot of second
(b) and first (c) band for a magnetic SL of $\delta$-function
barriers with $a = 4$, $b = 4$, and $P = 1$.} \label{fig5}
\end{center}
\end{figure}
%
%
\begin{figure}[ht]
\begin{center}
\includegraphics[width=10cm]{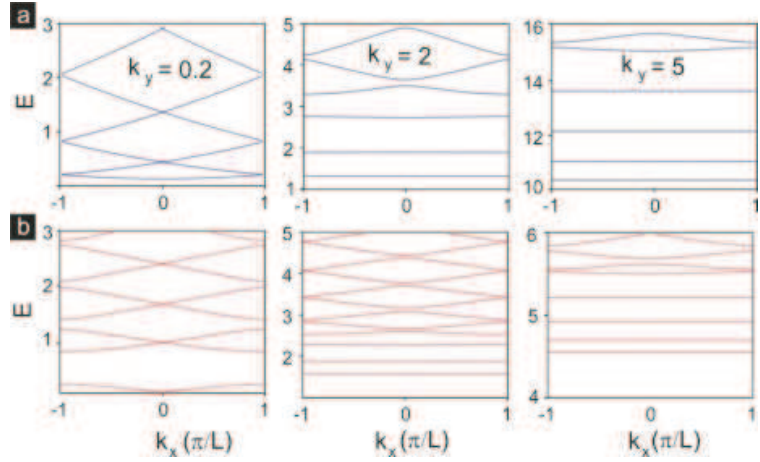}
\caption{ Dispersion relation  ($E$ vs $k$)  for a standard electron
in (a) and a Dirac electron in (b). The fixed values of $k_y$ are
shown in the panels and  $L=8$ and $P=1$ (the energy for standard
electron is measured in units of $\hbar \omega _{c}$ with $\omega
_{c} = \sqrt{eB_{0}/mc}$ and all distances in
$l_{B}=\sqrt{c\hbar/eB_{0}}$).} \label{fig6}
\end{center}
\end{figure}
\begin{figure}[ht]
\begin{center}
\includegraphics[width=10cm]{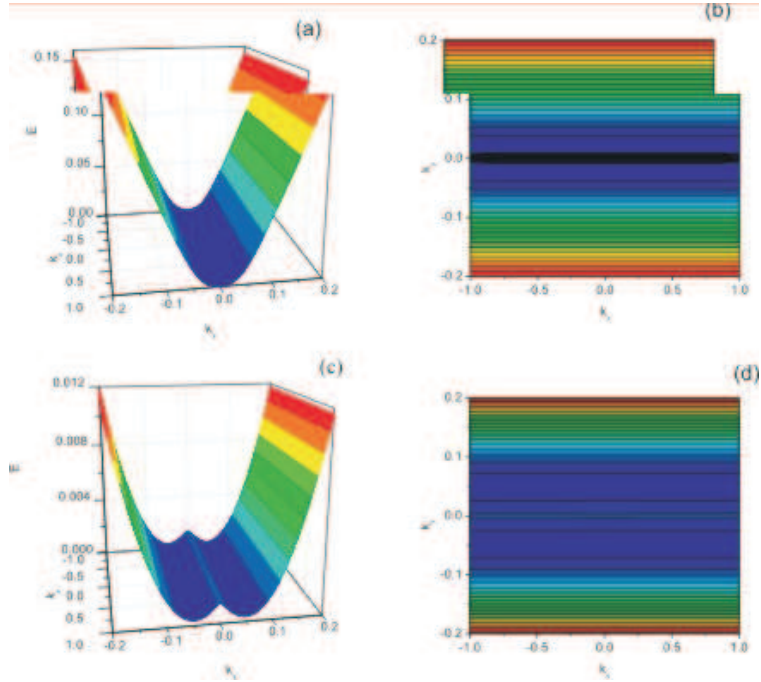}
\caption{First energy band for (a) standard and (c) Dirac electron
in SL of magnetic $\delta$-function barriers with $a = 100$, $b =
100$, and $P = 0.1$. (b) and (d) corresponding contour plots of (a)
and (c) (the energy for standard electron is measured in units of
$\hbar \omega _{c}$ with $\omega _{c} = \sqrt{eB_{0}/mc}$ and all
distances in $l_{B}=\sqrt{c\hbar/eB_{0}}$).} \label{fig7}
\end{center}
\end{figure}
\begin{figure}[ht]
\begin{center}
\includegraphics[width=10cm]{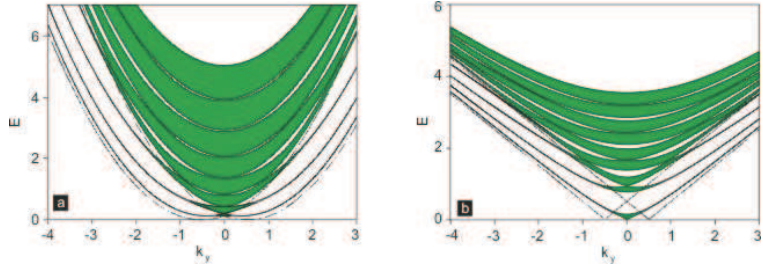}
\caption{Dispersion relation for a (a) standard electron and a (b)
Dirac electron. The period is $L= a+b = 8$ and the shaded (in green)
regions are the lowest six allowed bands. The solid curves in both
panels, the dash-dotted curves in (a) and the dashed ones in (b)
show bound states for free electron (the energy for standard
electron is measured in units of $\hbar \omega _{c}$ with $\omega
_{c} = \sqrt{eB_{0}/mc}$ and all distances in
$l_{B}=\sqrt{c\hbar/eB_{0}}$). }\label{fig8}
\end{center}
\end{figure}
\begin{figure}[ht]
\begin{center}
\includegraphics[width=10cm]{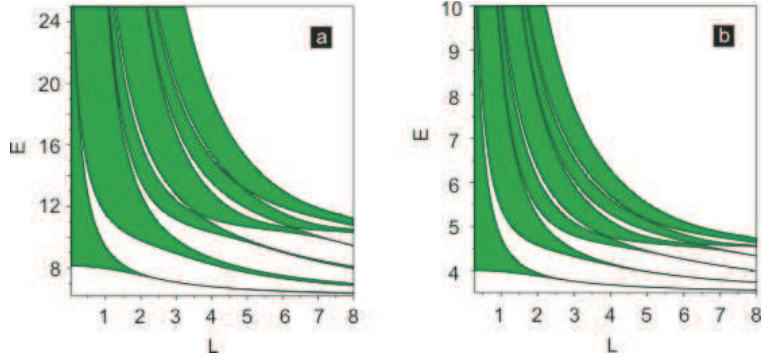}
\caption{Energy vs period $L = a+b$  for (a) standard electron and
(b) Dirac electron with fixed $k_{y} = 4$ (the energy for standard
electron is measured in units of $\hbar \omega _{c}$ with $\omega
_{c} = \sqrt{eB_{0}/mc}$ and all distances in
$l_{B}=\sqrt{c\hbar/eB_{0}}$). }\label{fig9}
\end{center}
\end{figure}
\begin{figure}[ht]
\begin{center}
\includegraphics[width=10cm]{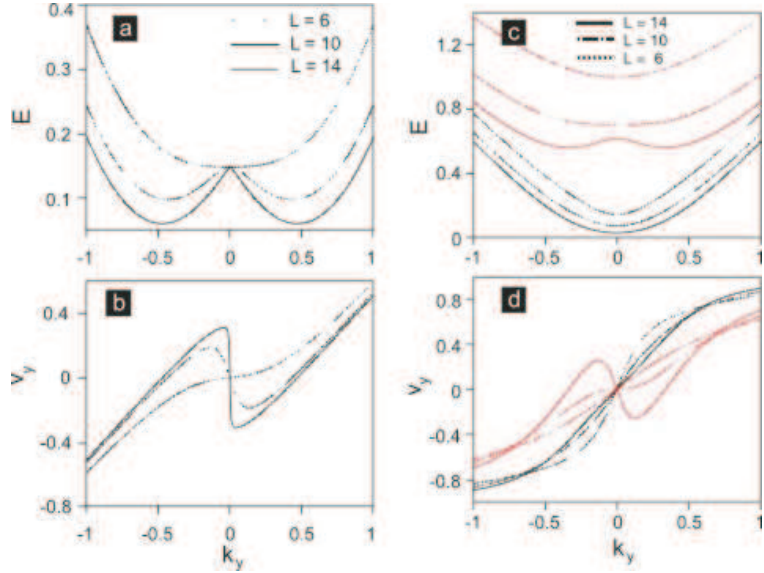}
\caption{(a) and (b): Lowest-energy band for a standard electrons
and drift velocity vs $k_{y}$ for three different values of $L$. (c)
and (d): First (black curves) and second (red curves) band for Dirac
electrons and the drift velocity vs $k_{y}$ for three different
values of $L$ (the energy for standard electron is measured in units
of $\hbar \omega _{c}$ with $\omega _{c} = \sqrt{eB_{0}/mc}$ and all
distances in $l_{B}=\sqrt{c\hbar/eB_{0}}$). }\label{fig10}
\end{center}
\end{figure}
\begin{eqnarray}
\label{E42}
\nonumber
 &\hspace*{-0.5cm} \displaystyle{M_{11} = s
  [\cos{\lambda^{+}_{2}}\cos{\lambda^{+}_{1}} - \sin\beta_{2}\sin\beta_{1}
  ]u_{N-1} - u_{N-2}}, \\
\nonumber & \hspace*{-0.6cm} \displaystyle{M_{12} = -is
  [\cos{\lambda^{+}_{2}}\sin\beta_{1} + \sin\beta_{2}\cos{\lambda^{-}_{1}}
  ]u_{N-1}},\\
\nonumber & \hspace*{-0.6cm} \displaystyle{M_{21} = -is
  [\cos{\lambda^{+}_{1}}\sin\beta_{2} + \sin\beta_{1}\cos{\lambda^{-}_{2}}
  ]u_{N-1} },\\
&\hspace*{-0.5cm}  \displaystyle{M_{22} = s
  [\cos{\lambda^{-}_{2}}\cos{\lambda^{-}_{1}} - \sin\beta_{2}\sin\beta_{1}
  ]u_{N-1} - u_{N-2}},
\end{eqnarray}
where $s=p_2p_1$,  $u_N\equiv u_N(\chi)$,  and
\begin{equation}\label{E43}
    \chi =  \displaystyle{\cos{(k_{1}a)}\cos{(k_{2}b)} - \left(\frac{k_{2}^2 +
    k_{1}^2+P^{2}}{2k_{2}k_{1}}\right)}\sin{(k_{1}a)}\sin{(k_{2}b)}.
\end{equation}
The reflection and transmission coefficients of the multi-unit
system are immediately obtained by substituting these expressions
into Eq. (\ref{E22}). The numerical results are shown in Figs.
\ref{fig3}, \ref{fig4} for finite superlattice with $N=10$ units.
Two different type of structures are considered as shown in the
insets to Figs. \ref{fig4}.
The transmission doesn't have $k_{y} \rightarrow -k_{y}$ symmetry
for the periodic system with magnetic delta up-down as is apparent
from Fig. \ref{fig4}(a). We contrast these results with the case in
which we used an arrangement of magnetic $delta$-function as in
previous structure plus another unit with opposite direction of
magnetic delta function. As is clearly shown in Fig. \ref{fig4}(b),
we have $k_{y} \rightarrow -k_{y}$ symmetry for the transmission
probability through this structure. The transmission resonances are
more pronounced, i.e., the dips become deeper, when the number of
barriers increases for both types of units.
But the gaps occur when the wave is mostly reflected. The position
of these gaps, which are especially pronounced as N increases, can
also be found from the structure of the Bloch phase $\zeta$, as
shown in Figs. \ref{fig3}(c) and (d).
In Fig. \ref{fig3} the bound states are shown by the blue solid
curves that are situated in the area $-k_{y}-P/2<E<-k_{y}+P/2$ in
case (a) and in $-k_{y}-P/2<E<-k_{y}$ for case (b) plus an area
located symmetric with respect to $k_{y}$. Notice in Fig.
\ref{fig3}(a) that several bound states merge into a resonant states
at $E = -k_{y} + P/2$.  This is different from Fig. \ref{fig2}(a)
where each bound state becomes a resonant state at $E=-k_{y}$.
\subsection{Spectrum of a superlattice}
Lets take $N\rightarrow \infty $. We can find the energy-momentum
relation from the previous standard calculation \cite{BHJ,Griff} by
using
\begin{equation}\label{E36d}
    \displaystyle{\cos(k_{x} L)= \frac{1}{2}\Tr \textbf{M}= \chi,}
\end{equation}
where $M$ is the characteristic matrix of one period, which results
into
\begin{equation}\label{Eq49s}
    \displaystyle{\cos{(k_{1}a)}\cos{(k_{2}b)} - \left(\frac{k_{2}^2 +
    k_{1}^2+P^{2}}{2k_{2}k_{1}}\right)\sin{(k_{1}a)}\sin{(k_{2}b)}}=\cos{k(b+a)}.
\end{equation}
With reference to the regions I and II shown in Fig. \ref{fig1}(a),
we write $k_{1} = [E^{2} - k_{y}^{2}]^{1/2}$ and $k_{2}= [E^{2} -
(k_{y} + P)^{2}]^{1/2}$ and show the solution for $E^{2}
> (k_{y} + P)^{2}$ in Fig. \ref{fig7}.
Differences of Eq. (\ref{Eq49s}) from the corresponding result of
Ref. \cite{ibra}, Eq. (\ref{E7}), for the case of Schr\"{o}dinger
electrons, is the term $P^2$ in the prefactor of the second term on
the right-hand side and the linear $E$ vs $k$ spectrum instead of
the quadratic one in Ref. \cite{ibra}. If $P$ is large the
differences become more pronounced. Our numerical results for the
energy spectrum are shown in Figs. {\ref{fig5}, \ref{fig6},
\ref{fig7}, \ref{fig8}, and \ref{fig9}. The results for standard and
Dirac electrons show similarities but also important differences.
The first band shows a qualitative difference near $k_y\approx 0$,
see Fig. \ref{fig7}. As Figs. \ref{fig5}, \ref{fig7}(a) and
\ref{fig7}(b) show, the band behavior in the $k=k_{x}$ direction for
fixed $k_{y}$ is constant and almost symmetric about $k_{y}=0$; the
motion becomes nearly 1D for relatively large $k_{y}$. From the
contour plots of Figs. \ref{fig7}(b) and (d), as well as from Fig.
\ref{fig5}(c), we infer a collimation along the $k_y$ direction,
i.e., $v_y\propto \p E/\p k_y \approx v_{F}$ and $v_x\approx0$,
which is similar to that found for a SL of electric potential
barriers \cite{Cheol} for some specific values of the barrier
heights. Also there are no gaps for $k_y\approx 0$ in Fig.
\ref{fig8}(a) but there are for the case of Dirac electrons as seen
in Fig. \ref{fig8}(b). This difference can be traced back to the
presence of $P^2$ in the dispersion relation Eq. (\ref{Eq49s}) when
compare to the same equation for the standard electron. The
even-number energy bands in Fig. \ref{fig9}(b)  are wider than those
in Fig. \ref{fig9}(a) and, as a function of the period, the energy
decreases faster for Dirac electrons. This behavior of the bands for
Dirac electrons is very similar to that for the frequency $\omega$
vs $k_y$ or $L$ in media with a periodically varying refractive
index \cite{yar}. This is clearly a consequence of the linear $E-k$
relation. Notice the differences between the lowest bands shown in
panels (a) and (b) in Fig. \ref{fig10} and in particular the
difference between the corresponding drift velocities as functions
of  $k_y$.
\section{A series of $\delta$-function vector potentials}
In the limit that the distance between the opposite directed
magnetic barriers decrease to zero the vector potential approaches a
$\delta$-function \cite{Castro}. We consider a series of magnetic
$\delta$-function vector potentials $A(x) =
\sum^{\infty}_{n=-\infty}{A_{0}\delta{(x - nL)}}$ as shown in Fig.
11(a). First we consider a single such potential, that is zero
everywhere except at $x=0$. We start with Eqs. (\ref{E6}) and
(\ref{E7}) which becomes now
\begin{eqnarray}
\hspace*{-1cm}  -i\left[
  d/d x + k_{y} + \ell_{B}A_{0}\delta(x) \right]\phi_{b} &=& E \phi_{a} , \label{E50}\\
\hspace*{-1cm}   -i\left[d/d x -k_{y} - \ell_{B}A_{0}\delta(x)
\right]\phi_{a} &=& E \phi_{b}.\label{E51}
\end{eqnarray}
The solutions are readily obtained  in the form
\begin{equation}\label{E52}
\hspace*{-1cm}     \phi_{a} = \left\{%
     \begin{array}{l}
               A \cos\left(\varepsilon x \cos{\theta} \right) + B \sin\left(\varepsilon x \cos{\theta} \right),\quad \quad \, \, \,  \quad \, x<0, \\
               C \cos\left(\varepsilon x \cos{\theta} \right) + D \sin\left(\varepsilon x \cos{\theta} \right),\quad \quad \, \, \,  \quad \, x>0,\\
             \end{array}
             \right.
\end{equation}
\begin{equation}\label{E53}
\hspace*{-1cm}     \phi_{b} = \left\{%
     \begin{array}{l}
               -i\left\{B\cos\left(\theta + \varepsilon x \cos{\theta}
    \right) - A \sin\left(\theta  + \varepsilon x \cos{\theta}\right)\right\},  \quad  x<0,  \\
               -i\left\{D\cos\left(\theta + \varepsilon x \cos{\theta}
    \right) - C \sin\left(\theta  + \varepsilon x \cos{\theta}\right)\right\}, \quad  x>0.\\
             \end{array}
             \right.
\end{equation}
Integrating   Eqs. (\ref{E50}), (\ref{E51}) around $0$ gives
\begin{eqnarray}\label{E54}
\hspace*{-1cm}  -i\int^{0+}_{0^{-}}\left[
  d/d x + (k_{y} + \ell_{B}A_{0}
  \delta(x)) \right]\phi_{b}\,dx &=& E\int^{0+}_{0^{-}} \phi_{a}\,dx , \\
 \hspace*{-1cm}  -i\int^{0+}_{0^{-}}\left[d/d x -(k_{y} + \ell_{B} A_{0}
  \delta(x)) \right]\phi_{a}\,dx &=& E\int^{0+}_{0^{-}}\phi_{b}\,dx
\end{eqnarray}
and
\begin{figure}[ht]
\begin{center}
\includegraphics[width=12cm]{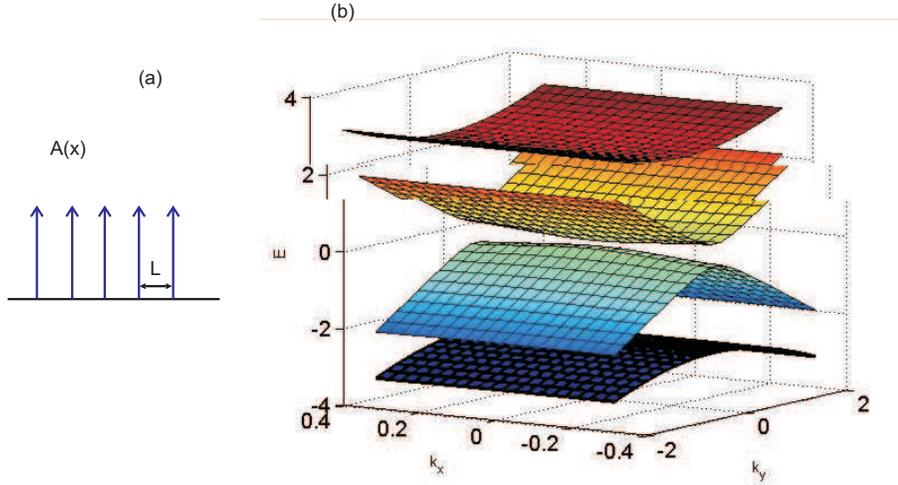}
\caption{(a) A  series of $\delta$-function vector potentials. (b)
Dispersion relation for the system shown in (a).}\label{fig12}
\end{center}
\end{figure}
\begin{equation}\label{E55}
    \phi_{b}(0^{-}) = \eta \phi_{b}(0^{+}), \quad\quad
    \phi_{a}(0^{+})= \eta \phi_{a}(0^{-}).
\end{equation}
We now consider the entire  series  of  $\delta$-function vector
potentials shown in Fig. \ref{fig12}(a) and use Eqs. (\ref{E54}) and
the periodic boundary condition
$\Psi_{I}(0)=e^{ik_{x}L}\Psi_{II}(L)$. The resulting dispersion
relation for the superlattice is
\begin{equation}\label{E56}
    \cos{(kL)} = \displaystyle{
    2\eta(1 + \eta^{2})^{-1}
    \cos(k_{x}L)},
\end{equation}
where $\eta = 1 + \ell_{B}A_{0}$.
From Eq. (\ref{E56}) we can find the energy spectrum as
\begin{equation}\label{E56a}
    E_{n}(k_{x},k_{y}) = \displaystyle{\pm \sqrt{2n\pi + k_{y}^{2} + \left(\frac{1}{L}\cos^{-1}\left(\frac{2\eta}{1 +
    \eta^2}\cos{(k_{x}L)}\right)\right)^2}};
\end{equation}
%
We can define
\begin{equation}\label{E57}
   \displaystyle{(1/L)\cos^{-1}\left[
     2\eta(1 + \eta^{2})^{-1} \cos(k_{x}L)\right]
    }=s,
\end{equation}
and obtain
\begin{equation}\label{E58}
    E = \pm [2n\pi + k^{2}_{y} + s^2]^{1/2}.
\end{equation}
The energy bands around the Dirac point are plotted in Fig.
\ref{fig10}(b). Notice that: 1) there is an opening of a gap at the
Dirac point, 2) the motion is strongly 1D, i.e. along the
$k_{y}$-direction, and 3) higher subbands have a smaller dispersion.
\section{Concluding remarks}
We developed a {\it magnetic} Kronig-Penney model for Dirac
electrons in graphene. The model is essentially a series of very
high and very narrow  {\it magnetic}  $\delta$-function barriers
alternating in signs. The treatment of the transmission through such
a  series of barriers followed closely the one developed in optics
for media in which the refractive index varies in space
\cite{born,yar}. We contrasted a few of the results with those for
standard electrons described by the Schr\"{o}dinger equation
\cite{ibra}.

In several cases the energy spectrum or the dispersion relation were
obtained analytically, cf. Eqs. (25), (39), (47), (48), and (57),
largely due to the simplicity of the model and the adapted method
from optics. For only two {\it magnetic}  $\delta$-function
barriers, opposite in sign, we saw several bound states, whose
number increases with $|k_y|$, and a reduction of the wavevector
range for which tunneling is possible, cf. Fig. 4(a). This is in
line with that reported earlier for single \cite{egg} and multiple
\cite{mass} barriers. The reduction becomes stronger as we increase
the number of barriers, cf. Fig. 4(b). We also made contact with
Snell's law in optics, cf. Eq. (36): the term $P/E$ represents the
deviation from this law.

An important feature of the superlattice results is a collimation of
an incident electron beam normal to the superlattice direction at
least for large wave vectors.  As easily seen from  Figs. 7 and 8,
for $|k_y|\geq 2$ we have $v_x\propto \p E/\p k_x \approx 0$ for the
first three minibands in the middle panels and nearly five minibands
in the right panels.  This occurs for both standard electrons and
Dirac  electrons but notice an important difference for
$|k_y|\approx 0$ shown clearly in Fig. 9. This collimation is
similar to that reported in Ref. \cite{Cheol} for superlattices
involving only electric barriers but with somewhat unrealistic large
barrier heights.

It is also worth emphasizing the differences and similarities in the
first two minibands and the corresponding drift velocities as
functions of $k_y$ for different periods $L$ and constant $k_x$ as
shown in Fig. 12. Notice in particular the resemblance between  the
drift velocities in the lowest miniband  for  standard  electrons
and the second miniband  for Dirac  electrons.

Given that ferromagnetic strips were successfully deposited on top
of a  2DEG in a semiconductor heterostructure \cite{nog}, we hope
they  will be deposited on graphene too and  that the results of
this paper will be tested in a near future.
\section*{ACKNOWLEDGEMENTS}
 \vspace*{-0.2cm}
We thank  Prof. A. Matulis for helpful discussions. This work was
supported by the Flemish Science Foundation (FWO-Vl), the Belgian
Science Policy
(IAP), the Brazilian National Research Council CNPq, and the Canadian NSERC Grant No. OGP0121756.\\


\section*{References}
\begin{thebibliography}{10}
\bibitem{no05}K.~S.~Novoselov, A.~K.~Geim, S.~V.~Morozov, D.~Jiang, M.I.~Katsnelson,
I.~V.~Grigorieva, S.~V.~Dubonos, and A.~A.~Firsov, Nature (London)
{\textbf {438}}, 197 (2005).

\bibitem{zh05}Y.~Zheng, Y.~W.~Tan, H.~L.~Stormer, and P.~Kim, Nature (London)
{\textbf {438}}, 201 (2005).

\bibitem{ka06}M.~I.~Katsnelson, K.~S.~Novoselov, and A.~K.~Geim, Nature
Physics {\textbf {2}}, 620 (2006); J. Milton Pereira Jr., P.
Vasilopoulos, and F. M. Peeters,  Appl. Phys. Lett. {\bf 90}, 132122
(2007).

\bibitem{pe93}F.~M.~Peeters and A.~Matulis, Phys.~Rev.~B {\textbf {48}},
15166 (1993).

\bibitem{Rei1}J. Reijniers, F.~M.~Peeters, and  A.~Matulis,
Phys.~Rev.~B {\bf {64}}, 245314 (2001).

\bibitem{Rei2}J. Reijniers, F.~M.~Peeters, and A.~Matulis,
Phys.~Rev.~B {\bf {59}}, 2817(1999).


\bibitem{egg}  A. De Martino, L. Dell' Anna, and R. Egger, Phys. Rev. Lett. {\bf {98}}, 066802
(2007).

\bibitem{rm08}M.~Ramezani Masir, P.~Vasilopoulos, A.~Matulis, and F.~M.~Peeters,
Phys.~Rev.~B {\textbf {77}}, 235443 (2008).

\bibitem{mass} M.~Ramezani Masir, P.~Vasilopoulos,  and F.~M.~Peeters, Appl. Phys. Lett. {\bf 93},  242103  (2008).

\bibitem{Park} S. Park and H. S. Sim, Phys. Rev. B {\bf {77}}, 075433 (2008).

\bibitem{Rakyta} L. Oroszlany, P. Rakyta, A. Kormanyos, C.J. Lambert, and J. Cserti,
Phys. Rev. B {\bf{77}}, 081403(R) (2008).

\bibitem{Ghosh} T. K. Ghosh, A. De Martino, W. H\"{a}usler, L. Dell'Anna, and R. Egger, Phys. Rev. B {\bf{77}}, 081404(R)
(2008).

\bibitem{zhai} F. Zhai and K. Chang, Phys. Rev. B {\bf{77}}, 113409
(2008).

\bibitem{MT}M. Tahir and K. Sabeeh, Phys. Rev. B {\bf{77}}, 195421 (2008).

\bibitem{Xu} Hengyi Xu, T. Heinzel, M. Evaldsson, and I.V. Zozoulenko, Phys. Rev. B {\bf{77}}, 245401
(2008).

\bibitem{Cheol} Cheol-Hwan Park, Feliciano Giustino, Marvin L. Cohen and Steven G.
Louie Nano Lett. {\bf{9}} , 1731, (2009).

\bibitem{Van} H. van Houten, B. J. van Wees, J. E. Mooij, C. W. J. Beenakker,
J. G. Williamson, and C. T. Foxon, Europhys. Lett. {\bf 5}, 721
(1988).

\bibitem{Spec} J. Spector, H. L. Stormer, K. W. Baldwin, L. N. Pfeiffer, and K.
W. West, Appl. Phys. Lett. {\bf 56}, 1290 (1990).

\bibitem{Siv}U. Sivan, M. Heiblum, C. P. Umbach, and H. Shtrikman, Phys. Rev.
B \textbf{41}, R7937 (1990).

\bibitem{Mole} L.W. Molenkamp, A. A. M. Staring, C.W. J. Beenakker, R. Eppenga, C.
E. Timmering, J. G. Williamson, C. J. P. M. Harmans, and C. T.
Foxon, Phys. Rev. B {\bf 41}, R1274 (1990).

\bibitem{Yac} A. Yacoby, M. Heiblum, V. Umansky, H. Shtrikman, and D. Mahalu, Phys. Rev. Lett. {\bf 73}, 3149
(1994).

\bibitem{mat} A. Matulis, F. M. Peeters, and P. Vasilopoulos, Phys. Rev. Lett. {\bf 72}, 1518 (1994).

\bibitem{nog} A. Nogaret, D. N. Lawton, D. K. Maude, J. C. Portal, and M. Henini, Phys. Rev. B {\bf 67}, 165317 (2003).

\bibitem{Castro}Vitor M. Pereira and Antonio H. Castro Neto,
arXiv: 0810.4539v3 (2009).

\bibitem{sharma}S. Ghosh and M. Sharma, arXiv: 0806.2951 (2008).

\bibitem{Dell}L. Dell'Anna and A. D. Martino, Phys. Rev. B {\bf 79}, 045420 (2009).

\bibitem{ibra} I. S. Ibrahim and F.~M.~Peeters, Phys. Rev. B {\bf 52}, 17321 (1995); Am. J. Phys. {\bf 63}, 171 (1995).

\bibitem{born} Max Born and Emil Wolf,  {\it Principles of Optics}, (Pergamon Press, 1980).

\bibitem{yar} Amnon Yariv, {\it Optical Waves in Crystals}, (John Wile \& Sons, Inc. 1976).

\bibitem{Ponomarenko} L. A. Ponomarenko, F. Schedin, M. I. Katsnelson, R. Yang, E. W. Hill,
K. S. Novoselov and A. K. Geim, Science {\bf 320}, 356 (2008).

\bibitem{Vadim} Vadim V. Cheianov, Vladimir Fa\'{l}ko and B. L. Altshuler, Science {\bf 315}, 1252 (2007);

\bibitem{BHJ} Bruce H. J. McKellar and G. J. Stephenson, Jr., Phys. Rev. C {\bf 35}, 2262
(1987).

\bibitem{Griff} David J. Griffiths and Carl A. Steinke. Am. J. Phys.  {\bf 69} , 137 (2001);
\end{thebibliography}
\end{document}